\documentclass[12pt]{article}
\usepackage{graphicx}

\newcommand\pubnumber{SNSN-323-63}
\newcommand\pubdate{\today}
\newcommand{\be}{\begin{equation}}
\newcommand{\ee}{\end{equation}}
\usepackage{amsfonts, amsmath, amsthm, amssymb}

\def\institute{$^{\star}$Univ. Grenoble Alpes, USMB, CNRS, LAPTh, F-74000 Annecy, France\\
$^{\dag}$CHEP, Indian Institute of
Science, Bangalore 560012, India \\ $^{\S}$University of Sussex, Brighton BN1 9RH, United Kingdom\\
$^{\ddagger}$Physical Research Laboratory, Navrangpura, Ahmedabad 380009, India}

\def\Title#1{\begin{center} {\Large #1 } \end{center}}
\def\Author#1{\begin{center}{ \sc #1} \end{center}}
\def\Address#1{\begin{center}{ \it #1} \end{center}}

\newcommand\pubblock{\rightline{\begin{tabular}{l} \pubnumber\\
         \pubdate  \end{tabular}}}
\newenvironment{Abstract}{\begin{quotation}  }{\end{quotation}}
\newenvironment{Presented}{\begin{quotation} \begin{center} 
             PRESENTED AT\end{center}\bigskip 
      \begin{center}\begin{large}}{\end{large}\end{center} \end{quotation}}
\def\Acknowledgements{\bigskip  \bigskip \begin{center} \begin{large}
             \bf ACKNOWLEDGEMENTS \end{large}\end{center}}

\def\beq{\begin{equation}}
\def\eeq#1{\label{#1}\end{equation}}
\def\eeqn{\end{equation}}

\def\beqa{\begin{eqnarray}}
\def\eeqa#1{\label{#1}\end{eqnarray}}
\def\eeqan{\end{eqnarray}}

\let\bar=\overbar

\def\Dslash{\not{\hbox{\kern-4pt $D$}}}
\def\dslash{\not{\hbox{\kern-2pt $\del$}}}

\def\ee{e^+e^-}

\def\msb{{\bar{\ssstyle M \kern -1pt S}}}

\begin{document}
\begin{titlepage}
\pubblock

\vfill
\Title{Probing CP nature of a mediator in associated production of dark matter with single top quark}
\vfill
\Author{Genevieve Belanger$^{\star}$, 
Rohini M. Godbole$^{\dag}$, Charanjit K. Khosa$^{\S}$\footnote{speaker}
and Saurabh D. Rindani$^{\ddagger}$}
\Address{\institute}
\vfill
\begin{Abstract}
We consider associated production of dark matter with single top quark,  in a simplified dark matter model with spin-0 mediators.
The produced top quark is polarized and the polarization depends on the CP of the mediator. We calculate both
the cross-section and top polarization for these processes. We compute angular asymmetries which  demonstrate the
 difference between  the polarization expected for the scalar or pseudoscalar mediator. Both the cross section 
 and top polarization are sensitive to  the CP property of the mediator, depending on the mediator mass. We find that these polarization 
 asymmetries add value to the determination of the CP property of the mediator particularly in the case of a state with indeterminate CP.
\end{Abstract}
\vfill
\begin{Presented}
$11^\mathrm{th}$ International Workshop on Top Quark Physics\\
Bad Neuenahr, Germany, September 16--21, 2018
\end{Presented}
\vfill
\end{titlepage}
\def\thefootnote{\fnsymbol{footnote}}
\setcounter{footnote}{0}
\section{Introduction}
Simplified dark matter models provide a convenient theoretical framework to interpret LHC results for dark 
matter(DM) searches. 
Associated production of a heavy quark pair with DM has been one of the important DM search channels from 
the very beginning of collider DM searches. However, it has been shown recently that associate production of single top 
with a pair of DM particles can also provide interesting reach\cite{Pinna:2017tay} and the experimental collaboration has 
 already started considering this channel\cite{CMS:2018vlx}. DM production in association with single top  has also been 
 studied in Ref. \cite{Plehn:2017bys}  in a simplified model 
 and in the context of two Higgs doublet model with an  additional pseudoscalar  in Ref. \cite{Pani:2017qyd}. 

We consider a simplified dark matter model with a spin-0 mediator ($\Phi$). 
The  Lagrangian we consider for the mediator couplings to the SM fermions and DM is :
\begin{equation}  {\cal {L}}_{\Phi} = g_{\chi} \Phi \bar \chi (\cos \theta+ i\sin \theta \gamma^5)  \chi +
\frac{g_v \Phi}{\sqrt{2}} \sum_{f=t,b} \frac{m_f}{v} \bar f (\cos \theta+ i\sin \theta \gamma^5) f
-\frac{1}{2} m_{\Phi}^2 \Phi^2 -m_{\chi} \bar\chi \chi. \label{eq1} \end{equation}
Here $\theta$ is the CP phase parameter  $\theta=$0 ($\frac{\pi}{2}$) represents pure scalar  (pseudoscalar) state
 and $v=174$ GeV. We consider the case where the  mediator has couplings only to the third generation fermions.

\begin{figure}[hbt]
\centering
\includegraphics[height=1.2in]{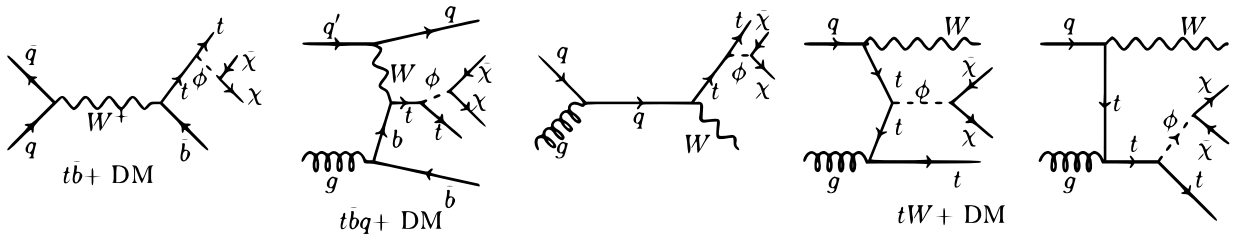}
\caption{Feynman diagrams for the dark matter production processes in association with single top quark }
\label{fig:feyndiag}
\end{figure}

In this scenario,  $t \bar t$ pair
production process in association with DM, had  been used to study the CP nature of the mediator
 \cite{Buckley:2015ctj,Haisch:2016gry}.  In this study the angular correlation between the two decay leptons has been exploited as an observable specific to the CP property.
We consider associated production of dark matter ($\chi$) with single top in a simplified dark matter model with a spin-0 mediator ($\Phi$) 
and  we investigate the effect of the CP nature of the mediator on the  cross-section  and on the top polarization.
 The production processes are shown in Figure ~\ref{fig:feyndiag}.
\section{Cross section and top polarization}
\begin{figure}[hbt]
\centering
\includegraphics[height=2.2in]{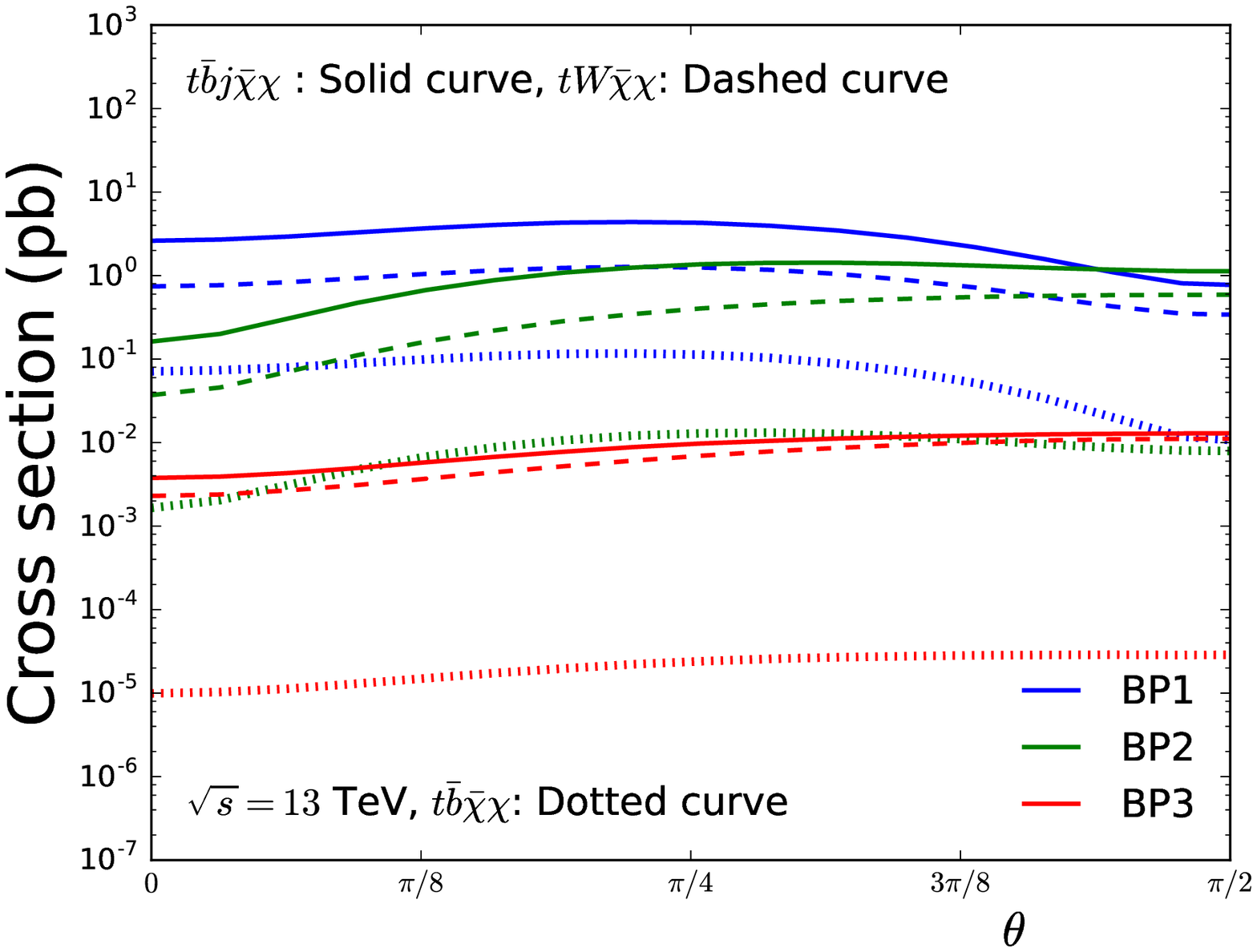}
\includegraphics[height=2.2in]{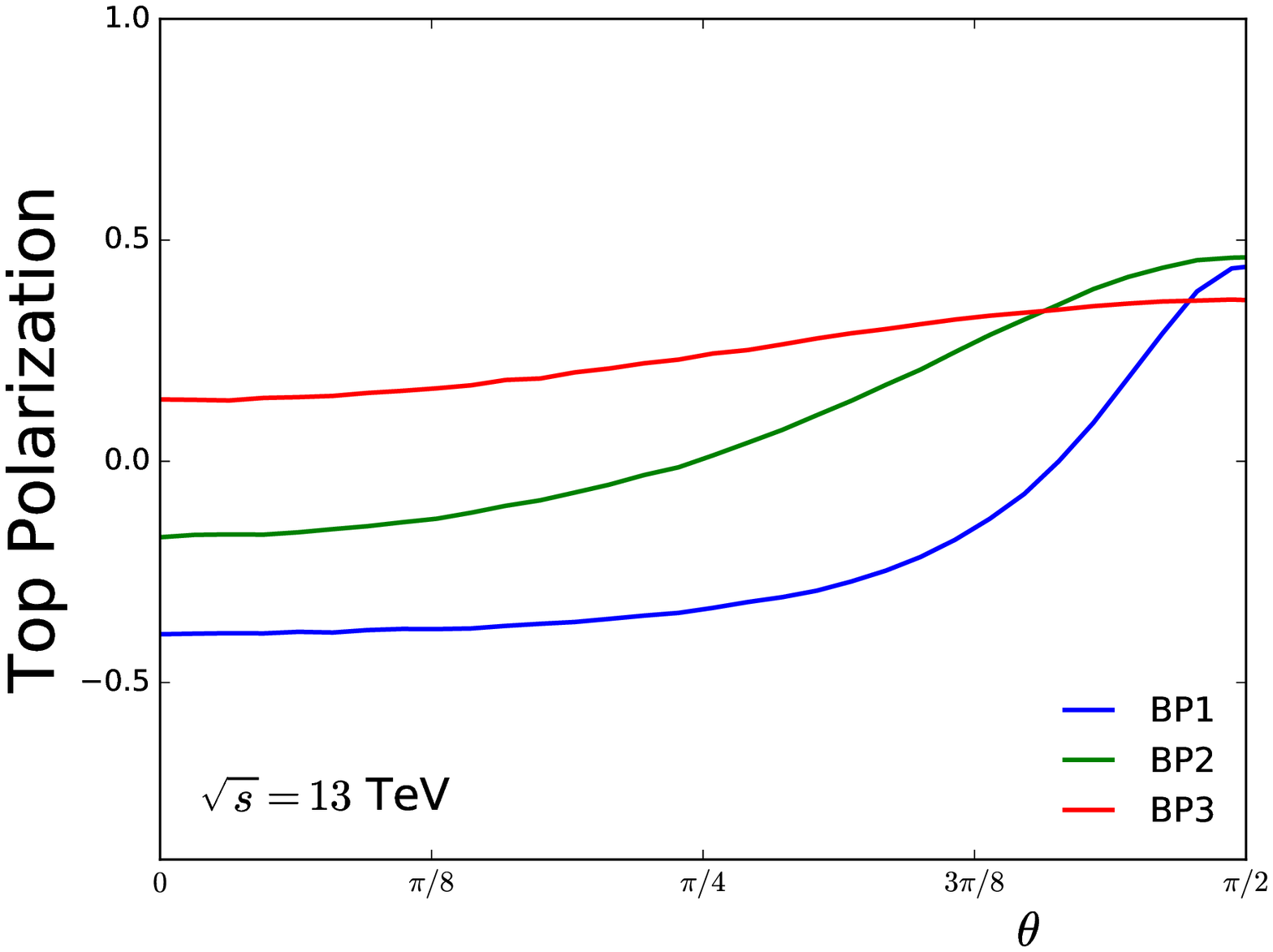}
\caption{Cross-section (including both $t$ and $\bar t$ processes) for single top in association 
with DM for the three benchmark points (left panel) and top polarization for $pp \rightarrow$ single top + DM processes (right panel), as a function of CP 
phase $\theta$}
\label{fig:crosspol}
\end{figure}
Model files are generated using FeynRules and cross-section is calculated using
 MadGraph 2.5.5 (pdf- NNPDF30.lo.as.130, lhaid-263000). We consider 5-flavour scheme (massive b quark) for all the process except t-channel
$ t \bar b j$  ( $\bar t  b j$) +DM process where 4 flavour scheme is considered. We have four model parameters
: $m_{\chi}, m_{\Phi}, g_{\chi}$, and $g_v$. We fix  $g_v$=1 and consider the  three benchmark
 points(BP)  given in Table~\ref{tab:blood}, for  other parameters. For these BPs, the upper bound on the relic density of DM is satisfied along with  the  DM direct detection constraints from DarkSide50\cite{Agnes:2018ves} and Xenon1T\cite{Aprile:2018dbl}. The cross-section for single top and DM processes as a function of $\theta$ is shown
 in Figure \ref{fig:crosspol} for each BP. Here we fix the mediator width to the scalar case value ($\theta=0$) for all values of $\theta$. We calculate the top polarization
\begin{equation} P=\frac{\sigma_+-\sigma_-}{\sigma_++\sigma_-}, \end{equation}
using helicity amplitudes in MadGraph, here $\sigma_+$ and $\sigma_-$ are the cross-sections for the positive and negative helicity top
 quarks. We can see in Figure \ref{fig:crosspol} that the top polarization and the cross-section have a different
 $\theta$ dependence. Hence polarization can provide additional discriminatory power.  Note in particular that for 
 the cross-section, the difference between BP1 and BP2 decreases at large values of $\theta$ whereas the polarization remains a good discriminant.
\section{Observables which reflect polarization}
Angular observables provide robust measures of polarization unaffected by physics beyond the standard model  in top decay (see e.g.\cite{Godbole:2006tq,Godbole:2018wfy}).
We compute the polar asymmetry 
\begin{equation} A_{\theta_l}=\frac{\sigma (\cos\theta_l > 0)-\sigma (\cos\theta_l< 0) }{\sigma (\cos\theta_l > 0)+\sigma (\cos\theta_l < 0)},  \end{equation}
where $\theta_l$ is an angle of the charged lepton (from top decay) with top direction of motion 
and the azimuthal asymmetry (about the top quark production plane) 
\begin{equation} A_{\phi}=\frac{\sigma (\cos\phi_l > 0)-\sigma (\cos\phi_l< 0) }{\sigma (\cos\phi_l > 0)+\sigma (\cos\phi_l < 0)}. \end{equation}
\begin{figure}[htb]
\centering
\includegraphics[height=2.2in]{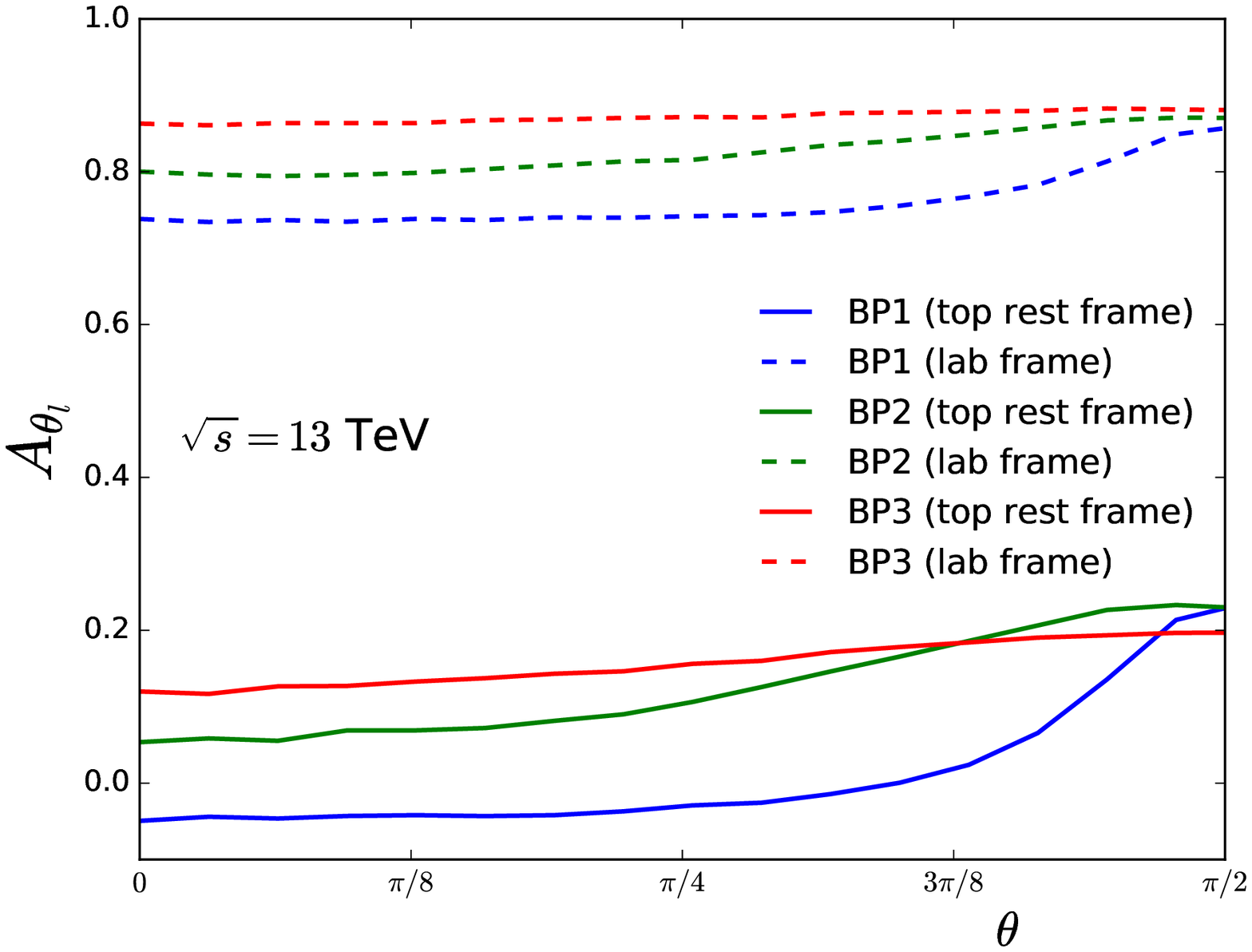}
\includegraphics[height=2.2in]{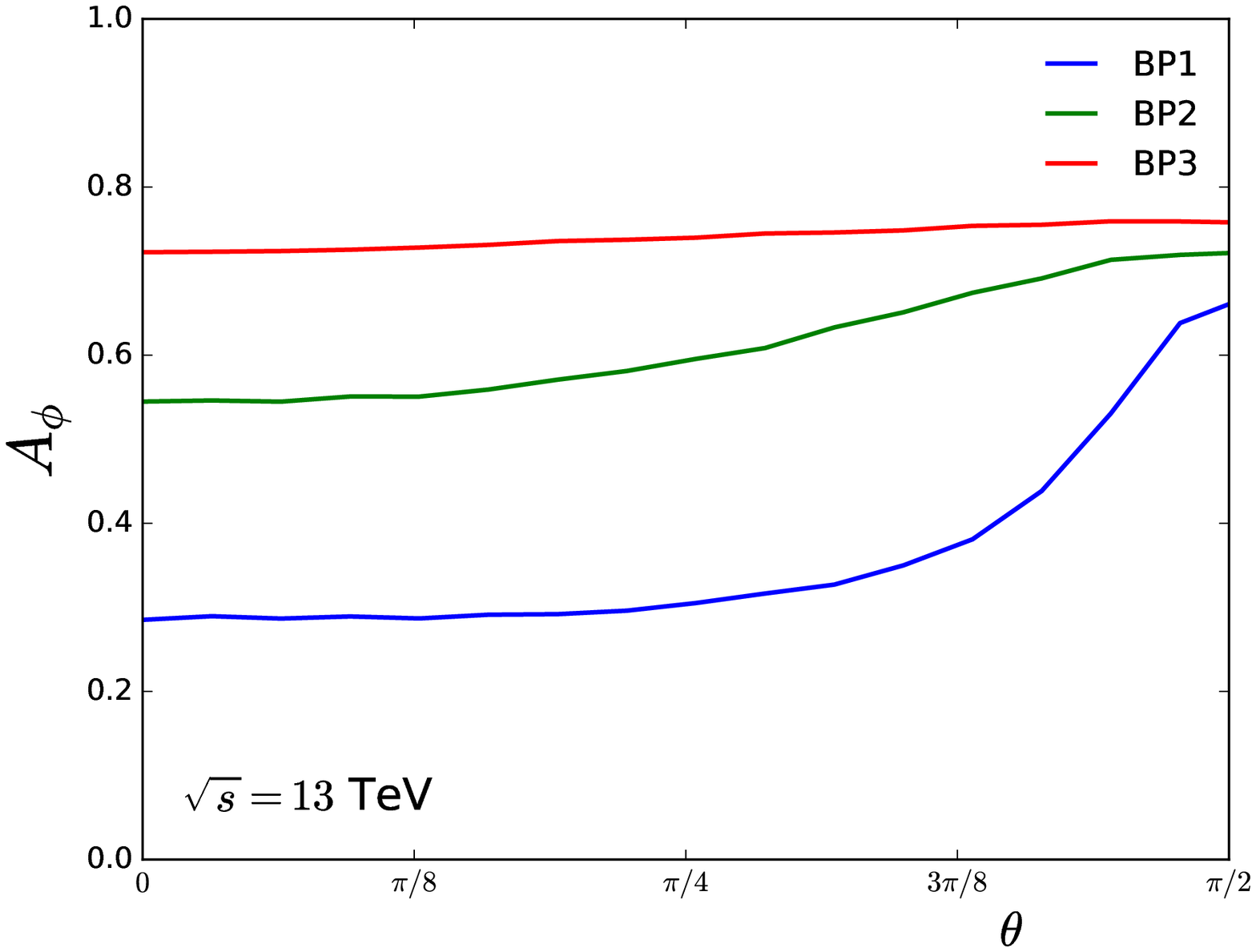}
\caption{ Charged lepton polar asymmetry (left panel) and azimuthal asymmetry for $pp \rightarrow$ single top + DM processes (right panel), 
as a function of CP phase $\theta$}
\label{fig:asym}
\end{figure}
We can see from Figure \ref{fig:asym} that $A_{\phi}$ has a strong dependence on $\theta$ when $\phi$ is light whereas the dependence of $A_{\theta_l}$ on the CP phase is milder.
Moreover note that $A_\phi$ can be a good discriminant between BP1 and BP2 for any value of $\theta$. 
\begin{table}[t]
\begin{center}
\begin{tabular}{l|ccc}  
 &  $m_{\phi}$ &  $m_{\chi}$ &  
$g_{\chi}$ \\ \hline
 BP1  &   10 GeV     &     4.5 GeV      &     0.35  \\
 BP2 &  100 GeV     &     49 GeV      &  0.5 \\ 
 BP3 &  400 GeV     &     180 GeV      &  1.0 \\\hline
\end{tabular}
\caption{Benchmark points for the model parameters}
\label{tab:blood}
\end{center}
\end{table}
\section{Summary and Conclusion}
\vspace{-.2cm}
We study the effect of the CP property of the mediator in simplified dark matter models. 
We consider associated production of DM with single top and calculate the
 cross-section as well as top polarization. 
 We find that the cross-section and top polarization have a different behaviour with respect to the CP phase $\theta$ thus offering complementary discriminatory power.
 Further study of the experimental sensitivity for these observables is in progress. 
\vspace{-.2cm}
\Acknowledgements
\vspace{-.2cm}
C.K.K. wishes to acknowledge support from the Royal Society-SERB Newton International Fellowship. 
 This work was supported in part by the
CNRS LIA-THEP  and the INFRE-HEPNET  of
CEFIPRA/IFCPAR (Indo-French Centre for the
Promotion of Advanced Research).

\end{document}